\documentclass{article}

\newcommand{\be}{\begin{equation}}
\newcommand{\ee}{  \end{equation}}
\newcommand{\ba}{\begin{eqnarray}}
\newcommand{\ea}{  \end{eqnarray}}

\begin{document}

\title{Nuclear Masses, Chaos, and the Residual Interaction} 

\author{A. Molinari \\ Dipartimento di Fisica Teorica dell' Universita
di Torino \\
and Instituto Nationale di Fisica Nucleare, Sezione di Torino, \\
Torino, Italy \\ H. A. Weidenm\"uller \\ Max--Planck--Institut f\"ur
Kernphysik, Heidelberg, Germany}

\maketitle

\noindent
Corresponding author: Hans A. Weidenm\"uller, Max-Planck-Institut
f\"ur \\ Kernphysik, P. O. Box 102980, 69029 Heidelberg, Germany\\
Telephone: 49/6221/516291 FAX: 49/6221/516602\\
email: Hans.Weidenmueller@mpi-hd.mpg.de        

\begin{abstract}
We interpret the discrepancy between semiempirical nuclear mass formulas
and actual nuclear masses in terms of the residual interaction. We show
that correlations exist among all binding energies and all separation
energies throughout the valley of stability. We relate our approach to
chaotic motion in nuclei.
\end{abstract}


\noindent
Pacs: 02.50.-r, 05.40.-a, 75.10.Nr

{\it Motivation.}
The masses of atomic nuclei are keys to the understanding of many
physical and astrophysical processes. For this reason it is important
to construct reliable theoretical models for the values of the nuclear
masses or, equivalently, of the binding energy $B(A)$ as function of
mass number $A$. (For simplicity of notation we suppress the additional
dependence on proton number $Z$). This function is also needed to
predict the masses of (as yet) unknown nuclei.

The standard approach to a global modelling of the function $B(A)$
starts out from the liquid--drop model for the nucleus and considers in
addition shell corrections as well as corrections due to pairing. The
resulting semi--empirical mass formula contains about 30 parameters and
is fitted to a large number of data. Years of painstaking work have
culminated in a best fit~\cite{moel} which reproduces the data points
very well albeit not exactly. The overall difference between that best
fit and the actual data is of the order of 0.5 Mev, with a tendency to
become smaller for heavier nuclei. That figure, although very small in
comparison with the total binding energy, is not negligible. Other
approaches~\cite{sam,duf} lead to similar differences.

Bohigas and Leboeuf~\cite{boh} have suggested that the discrepancy can
be attributed to the fact that the nuclear dynamics is partly chaotic.
Using the semiclassical approximation and Berry's result~\cite{berry} for
the form factor of the two--point correlation function, these authors
have estimated the contribution to the variance of the binding energy due
to chaotic single--particle motion. For the square root of the variance,
the estimate yields
\be
\sigma = \frac{2.78}{A^{1/3}} \ {\rm MeV} \ .
\label{boh1}
\ee
Like the shell--correction employed in the best--fit procedure of
Ref.~\cite{moel}, Eq.~(\ref{boh1}) is derived in the framework of the
single--particle picture and is, thus, based upon a mean--field
description. However, the authors of Ref.~\cite{boh} state that this
equation ``presumably gives an estimate of the mass fluctuations from
neglected many--body effects''. When plotted versus mass number, the
result~(\ref{boh1}) is in good agreement with the observed discrepancy
except for the lightest nuclei, see Fig.~1 of Ref.~\cite{boh}. 

The work of Ref.~\cite{boh} has led to a lively discussion in the
community. In this letter, we wish to address some of the points that
have come up in the discussion. Our point of view differs from that of
Ref.~\cite{boh} in that we interpret the discrepancy in terms of
many--body effects due to the residual interaction of the shell model
and explore the consequences of such an interpretation. We are primarily
interested in the conceptual implications of such an approach.
Many--body effects have been addressed in a number of publications, see,
for instance, Refs.~\cite{hir1,hir2,bar,vel}. We discuss some of these
works below.

{\it The Two--Body Random Ensemble.}
At the outset, we have to specify the residual interaction. We assume
that the residual interaction is of two--body form. We do so only to
have a specific model within which we can work. We are aware of the
fact that there is evidence for weak three--body interactions in
nuclei~\cite{Pi01}. It will be seen that allowing for three--body
forces will not affect our conclusions.

Within every major shell, the residual two--body interaction is specified
in terms of a set of two--body matrix elements. All possible two--body
interactions are taken into account by considering the two--body matrix
elements as Gaussian--distributed random variables. This defines the
two--body random ensemble (TBRE) of the shell model~\cite{fren,boh3}.
The TBRE has the advantage of leading to generic statements (valid for
almost all two--body interactions). This is useful since not all of the
two--body matrix elements of the residual interaction of the shell model
are precisely known either empirically or theoretically, especially in
medium--weight and heavy nuclei, and for higher--lying shells in any
nucleus. Moreover, the use of the TBRE allows us to estimate variances of
observables and, thus, the range within which values of the observables
will vary as we vary the parameters of the residual interaction or,
equivalently, to estimate the uncertainty of the observables due to
incomplete knowledge of the residual interaction.

If one applies the TBRE to the calculation of binding energies (more
precisely: to estimate corrections to the semi--empirical mass formula as
employed in Ref.~\cite{moel}), the binding energies will vary from one
realization of the TBRE to the next. Technically speaking, the binding
energies become random variables. Again, their variances express our
uncertainty about the residual interaction. We use these variances as a
measure of the deviations from empirical values of the binding energies
that are expected to occur in semi--empirical mass formulas which neglect
parts or all of the residual interaction. (The actual formula used in
Ref.~\cite{moel} does not disregard the residual interaction altogether
but does take into account the pairing force and the symmetry energy.
The version of the TBRE used here must take this fact into account by
keeping fixed the corresponding two--body matrix elements, and by
considering as random only the remainder).

{\it Binding Energies and Separation Energies.}
We discuss the application of the TBRE to binding energies. We first
recall the connection between nuclear binding energies and nucleon
separation energies. We denote the separation energy of the least--bound
nucleon in nucleus $A$ (which is either a neutron or a proton) by $S(A)$,
again omitting for simplicity the dependence of $S$ on $Z$. We can
completely decompose the nucleus $A$ into its constituent nucleons by
removing one nucleon after the other. Thus we have the identity
\be
B(A) = \sum_{j = 2}^A S(j) \ .
\label{1}
\ee
Likewise, since after removal of $k$ nucleons we reach the nucleus with
mass number $A - k$ (with $k$ integer), we also have
\be
B(A) = B(A - k) + \sum_{j = A - k + 1}^A S(j) \ .
\label{2}
\ee
Given the initial nucleus with mass number $A$ and proton number $Z$,
the $S(j)$ in Eqs.~(\ref{1}) and (\ref{2}) are uniquely specified by the
requirement that we always remove the least--bound nucleon. With this
understanding, Eqs.~(\ref{1}) and (\ref{2}) hold for all nuclei (and not
only for those in the bottom of the valley of stability) and for all values
of $k$. More equations of the same type are obtained when we allow for the
removal of the least--bound proton or least--bound neutron when that is not
the least--bound nucleon. Hence, equations of the type of Eqs.~(\ref{1})
and (\ref{2}) constitute a dense network of identities that hold throughout
the mass valley.

It was stated above that as we apply the TBRE to calculate corrections to
the binding energies predicted by the model of Ref.~\cite{moel}, such
binding energies become random variables. The same holds for the nucleon\
separation energies. For clarity these are denoted by the letters ${\cal
B}$ and ${\cal S}$, respectively. The random variables ${\cal B}(A)$ and
${\cal S}(j)$ must obey equations of the type~(\ref{1}) and (\ref{2}),
since by the very definition of the binding energy and the separation
energy, every realization of ${\cal B}(A)$ and ${\cal S}(j)$ in the
framework of the TBRE must obey these equations. Thus,
\be
{\cal B}(A) = \sum_{j = 2}^A {\cal S}(j) \ ,
\label{3}
\ee
\be
{\cal B}(A) = \sum_{j = A - k + 1}^A {\cal S}(j) + {\cal B}(A - k) \ .
\label{4}
\ee

{\it Correlations.}
We recall that the deviations of the data from the empirical mass formula
decrease with increasing values of mass number $A$, see Eq.~(\ref{boh1}).
We now show that this fact combined with Eqs.~(\ref{3}) and (\ref{4})
implies strong correlations between the random variables ${\cal B}(A)$
and ${\cal S}(A)$ for all values of $A$. We recall that stochastic
contributions to these variables arise from the TBRE. If we assume that
all variables ${\cal B}(A)$ and ${\cal S}(A)$ are uncorrelated, it
follows from Eq.~(\ref{3}) that
\be
{\rm Var} \big( {\cal B}(A) \big) = \sum_{j = 2}^A {\rm Var} \big(
{\cal S}(j) \big) \ .
\label{5}
\ee
The variance is, by definition, positive semidefinite. Therefore,
Eq.~(\ref{5}) implies that ${\rm Var} \big( {\cal B}(A) \big)$ is a
non--decreasing function of $A$. (That same conclusion was drawn already
in Ref.~\cite{mol} albeit in the framework of a specific model. Here we
establish it in full generality). This result is in striking contrast to
the monotonic decrease of the discrepancy with increasing mass number
displayed by the data and by Eq.~(\ref{boh1}). We conclude that there
exist strong correlations between binding energies and separation
energies.

Allowing for such correlations to exist, we arrive at a modified form of
Eq.~(\ref{5}). With the covariance of two random variables $X$ and $Y$
defined as ${\rm Cov} \big( X Y \big) = \overline{ X Y } - \overline{X} \
\overline{Y}$, we have
\be
{\rm Var} \big( {\cal B}(A) \big) = \sum_{j = 2}^A {\rm Var} \big(
{\cal S}(j) \big) + \sum_{i \neq j; i,j = 2}^A {\rm Cov} \big( {\cal S}(i)
{\cal S}(j) \big) \ .
\label{6}
\ee
In order to attain the observed monotonic decrease of the discrepancy
with mass number, the sum of the covariances must be negative semidefinite
for all values of $A$ and cannot vanish identically. This definitely
implies the existence of correlations among the ${\cal S}(j)$'s. Given
this fact, we expect correlations to exist also among binding energies
pertaining to different values of $A$. Indeed, Eq.~(\ref{3}) implies
\be
{\rm Cov} \big( {\cal B}(A) {\cal B}(A-k) \big) = \sum_{j = 2}^{A-k}
{\rm Var} \big( {\cal S}(j) \big) + \sum_{j_1 = 2}^A \sum_{j_2 = 2}^{A-k}
\big( 1 - \delta_{j_1 j_2} \big) {\rm Cov} \big( {\cal S}(j_1) {\cal S}
(j_2) \big) \ . 
\label{7}
\ee
This expression is, in general, different from zero. In contrast to the
case of the separation energies we have, however, no firm empirical
evidence in the present framework that these correlations do indeed exist.
They may vanish accidentally for all values of $A$. Even in this case we
expect correlations of higher order than the second to differ from zero.
Indeed, it takes a very special form of the probability density of random
variables connected by linear relations like Eq.~(\ref{3}) to be
uncorrelated. We conclude that within the framework of the TBRE, all
binding energies and all separation energies throughout the mass valley
are correlated.

{\it Discussion.}
From the point of view of nuclear physics, this result is not terribly
surprising. It mirrors the fact that a change of the residual interaction
within a major shell will affect the binding energies and the separation
energies of all nuclei pertaining to that shell. As we cross the boundary in
mass number separating major shells, the dominant two--body matrix elements of
the residual interaction change, and we might expect a weakening of the
correlations. Our result shows no trace of such an effect, and this is
probably its most surprising feature from the nuclear physics point of view.

Unfortunately, we cannot draw any conclusions about the range of the
correlations. Still, the following model is instructive. We assume that the
correlations between the separation energies are minimal and restricted to
nearest neighbors only so that ${\rm Cov}( {\cal S}(j_1) {\cal S}(j_2) )
\propto \delta_{j_1, j_2 \pm 1}$. For the range of the correlations, this is a
worst--case scenario. Our assumption is consistent with Eq.~(\ref{6}) provided
all these covariances are negative definite. By a suitable choice of the
values of the covariances, it is possible to account for the monotonic
decrease of $\sigma$ with $A$ as postulated in Eq.~(\ref{boh1}). However, even
that model implies long--range correlations between the binding
energies. Indeed, under the assumptions of the model, the right--hand side of
Eq.~(\ref{7}) becomes equal to ${\rm Var}( {\cal B}(A-k)) + {\rm Cov}( {\cal
S}(A-k+1) {\cal S}(A-k)$. Thus, the covariances of ${\cal B}(A-k)$ with ${\cal
B}(A)$ have for all values of $A$ the same values! We have not been able to
design a model which would satisfy Eqs.~(\ref{6}) and (\ref{7}) with
short--range correlations between all ${\cal S}$'s and all ${\cal B}$'s. From
the point of view of the TBRE, we expect the correlations to extend over a
range of mass numbers given by the range of major shells, see
Ref.~\cite{pap1}.

Statistical fluctuations of nuclear spectra are commonly described not in
the framework of the TBRE but rather in the framework of the Gaussian
orthogonal ensemble of random matrices (GOE). From the point of view of
canonical random--matrix theory, the existence of long--range correlations
is unexpected. Indeed, at present there exists no conceptual framework
within the GOE in which the existence of such correlations could be
accommodated. Thus, the data point to a further limitation of canonical
random--matrix theory. Earlier work on $sd$--shell nuclei~\cite{pap1} has
revealed such limitations in a more restricted framework by proving the
existence of correlations between levels in the same nucleus carrying
different quantum numbers and between levels in different nuclei belonging
to the same major shell.

What is the relation between our point of view -- which blames the
discrepancy between data and the nuclear mass formula on the residual
interaction -- and that taken in Ref.~\cite{boh} -- which puts the blame
on order--chaos coexistence? We identify chaos in nuclei with the
existence of spectral fluctuations of the Wigner--Dyson type. This
statement applies to levels of fixed spin and parity within the same
nucleus. In the following three paragraphs we present arguments to the
effect that such spectral fluctuations are always due to the residual
interaction. Given this fact, we conclude that the two points of view
are not fundamentally different especially in view of the
remark~\cite{boh} that Eq.~(\ref{boh1}) ``presumably gives an estimate
of the mass fluctuations from neglected many--body effects''.

To justify our view that the residual interaction is the agent responsible
for chaos in nuclei, we consider first spherical nuclei. Here the
single--particle motion is regular and does not lead to level sequences
which obey Dyson--Mehta statistics. This changes when the residual
interaction is included. Extensive numerical studies by many authors,
mainly of nuclei in the $sd$--shell, have shown that in this case,
nearest--neighbor spacing distribution and Mehta--Dyson statistic do
agree with GOE predictions whenever the number of nucleons in the shell
exceeds three or four. This is true provided the matrix elements of the
residual interaction are sufficiently strong to mix configurations
pertaining to different single--particle energies within the same major
shell. For a review, see Ref.~\cite{zel}. Recent theoretical studies of
the TBRE confirm this picture for the $sd$--shell~\cite{pap} and show
that almost all two--body residual interactions cause chaos. Generalizing
this argument we are led to expect that in every major spherical shell,
a sufficiently strong residual interaction generically produces chaos.
(This statement is not at variance with the existence of regular spectral
features near the ground state or like the giant resonances. It relates
to spectral statistics, i.e., to the joint properties of a large number
of levels). We add in parenthesis that the Skyrme--Hartree--Fock
approach~\cite{ber} or other similar approaches take account of
many--body effects in terms of an effective single--particle density
functional. In this approach the line separating single--particle and
many--particle effects becomes blurred.

The situation is somewhat different in deformed nuclei. For sufficiently
strong deformations, the single--particle motion alone may be chaotic,
and the single--particle spectrum may obey Wigner--Dyson statistics. But
strong deformations occur in the middle between large major shells, so
we deal with many valence nucleons. And when we put several or many
nucleons into these deformed single--particle states and observe the
exclusion principle, we generate a spectrum which by construction is a
superposition of several or many GOE spectra. It is known that such
superpositions tend rapidly (with increasing nucleon number) towards
the Poisson spectrum which is characteristic of integrable motion. This
suggests that also in deformed nuclei, the residual interaction is
needed to mix the configurations and to generate spectral properties of
the Wigner--Dyson type. We are not aware, however, of studies to show
under which conditions sufficiently strong mixing is attained. 

A different view of chaos in nuclei is offered by the Interacting Boson
Model which covers the case of deformed nuclei but also that of nuclei
with vibrational spectra etc. That is an effective model which describes
the low--energy spectra of nuclei with many nucleons outside of closed
shells, i.e., of nuclei where shell--model calculations are ineffective,
in terms of several interacting bosons of angular momenta zero and two.
Without any residual interaction between nucleons, the model would not
exist. Studies of chaos within this model~\cite{whel} have shown that
chaos is generic except near those parameter values where dynamical
symmetries persist.

The available evidence then suggests that except for very special cases
where symmetries dominate, it is legitimate to say that almost all forms
of the residual interaction cause chaos, and that chaos would not exist
without the residual interaction. This view is also supported by the
empirical evidence, i.e., the agreement of the spectral fluctuations of
the nuclear data ensemble with Wigner--Dyson statistics~\cite{boh2},
and the available evidence at lower excitation energies, see
Ref.~\cite{abul} and references therein.

This view sheds new light on attempts to reduce the discrepancy between
mass formulas and data by taking into account the residual interaction
in some way, either by explicit diagonalization, or by using relations
between binding energies of the Garvey--Kelson type or some other local
(in $A$) information. Several successful such attempts have been
published, see, for instance, Refs.~\cite{hir1,hir2,bar,vel}. Inasmuch
as the residual interaction produces chaos, this success {\it does not
contradict} the interpretation of the discrepancy in terms of chaotic
nuclear motion. The same remark would apply to versions of the
energy--density functional theory where the residual interaction is
incorporated in the single--particle Hamiltonian.

In some papers~\cite{aberg,bar} chaos in nuclei is interpreted as
limiting the accuracy with which spectra or binding energies in nuclei
can be calculated. It is claimed that the error estimated in Eq.~(\ref{boh1})
constitutes a bound on the possible accuracy of dynamical calculations of
nuclear masses, and any reduction of this bound is then taken as proof
that the discrepancy does not have its origin in chaos. We disagree with
this point of view. The spectra of nuclei in the $sd$--shell, for instance,
do display chaos but can be calculated without any limitation on the
numerical accuracy. This point is discussed in Ref.~\cite{pap}. More
generally, it was shown in Ref.~\cite{cas} that while classical chaos
does indeed limit dynamical calculations to a time scale given by the
Lyapunov coefficient, this is not true in general for quantum calculations.

In a recent paper~\cite{boh4} it was shown that correlations among
binding energies can also be accounted for in the framework of the
mean--field approach. So far, this work is restricted to neighboring
isotopes differing by a few units in mass number.

In summary, we have interpreted the discrepancy between the nuclear
mass formula and nuclear masses in terms of effects due to the residual
interaction. Using the two--body random ensemble, we have shown that
correlations exist between all binding energies and all separation
energies throughout the valley of stability. To the best of our knowledge,
this is the first evidence for such wide--ranging correlations in nuclei.
We have argued that chaos in nuclei is generically caused by the residual
interaction and does not exist without it. Therefore, we believe that
there is no substantial difference between our view and the one taken in
Ref.~\cite{boh}. There are successful attempts to reduce the discrepancy
by taking into account parts of the residual interaction not contained in
the semiempirical mass formula in theoretical calculations. Such attempts
do not contradict the interpretation of the discrepancy displayed in
Ref.~\cite{boh} in terms of chaotic nuclear motion.

The authors are grateful to T. Papenbrock for helpful discussions.

\end{document}